\documentclass[prb,twocolumn,showpacs,amsmath,amssymb]{revtex4}
\usepackage{graphicx}
\usepackage{amssymb}
\usepackage{bm}
\usepackage{epstopdf}

\newcommand{\be}{\begin{equation}}
\newcommand{\ee}{\end{equation}}
\newcommand{\bea}{\begin{eqnarray}}
\newcommand{\eea}{\end{eqnarray}}
\newcommand{\besa}{\begin{subeqnarray}}
\newcommand{\eesa}{\end{subeqnarray}}
\newcommand{\bean}{\begin{eqnarray*}}
\newcommand{\eean}{\end{eqnarray*}}

\begin{document}

\title{Dipoles on a Two-leg Ladder}

\author{S{\o}ren Gammelmark}
\affiliation{Department of Physics and Astronomy, Aarhus University, DK-8000 Aarhus C, Denmark}
\author{Nikolaj Thomas Zinner}
\affiliation{Department of Physics and Astronomy, Aarhus University, DK-8000 Aarhus C, Denmark}

\date{\today}

\begin{abstract}
We study polar molecules with long-range dipole-dipole interactions 
confined to move on a two-leg ladder for different orientations of the 
molecular dipole moments with respect to the ladder. Matrix product states are employed to 
calculate the many-body ground state of the system as function of lattice filling fractions, 
perpendicular hopping between the legs, and dipole interaction strength. We show that the 
system exhibits zig-zag ordering when the dipolar interactions are predominantly repulsive. 
As a function of dipole moment orientation with respect to the ladder, we find that there is 
a critical angle at which ordering disappears. This angle is slightly larger than the 
angle at which the dipoles are non-interacting along a single leg.
This behavior should be observable using current experimental techniques.
\end{abstract}


\maketitle

\section{Introduction}
Cold and ultracold molecules have recently become experimentally 
accessible
\cite{ospelkaus2008,ni2008,deiglmayr2008,lang2008,ospelkaus2010,ni2010,shuman2010,miranda2011,chotia2012}
and such systems hold great promise for exploration of quantum systems with 
long-range and anisotropic interactions with applications to quantum 
simulation, quantum metrology, and quantum information and computation \cite{baranov2008,lahaye2009,carr2009,baranov2012}.
However, strong dipolar interactions can induce collapse instabilities \cite{lushnikov2002} 
as well as strong chemical reaction losses \cite{ospelkaus2010}. These problems can be 
overcome in low-dimensional geometries where the dipolar particles are confined to 
either two-dimensional (2D) planes or one-dimensional (1D) tubes with and without the 
presence of lattice potentials.
A number of interesting predictions have been made for the phases of 
system with dipolar interactions, 
including exotic superfluids \cite{bruun2008,cooper2009,fellows2011,barbara2011,silva2013},
Luttinger liquids \cite{citro2007,citro2008,kollath2008,chang2009,huang2009,dalmonte2010}, Mott insulators 
\cite{arguelles2007,bauer2012}, interlayer 
pairing \cite{wang2007,pikovski2010,potter2010,zinner2012a}, non-trivial quantum critical points \cite{leche2012,tsvelik2012},
modified confinement-induced resonances \cite{sinha2007,bartolo2013,deu2013,guan2013}, 
roton modes and stripe instabilities \cite{yamaguchi2010,sun2010,zinner2011,ticknor2011,hufnagl2011,babadi2011,macia2012,kusk2012}, and 
crystallization \cite{baranov2005,buchler2007,mora2007,lu2008,dalmonte2011,matveeva2012,abedin2012,armstrong2012,knap2012}, as
well as formation of chain complexes \cite{wang2006,shih2009,klawunn2010,deu2010,santos2010,wunsch2011,volosniev2012,zinner2012b,volosniev2013}.

In the present paper we consider dipolar particles confined to move in an 
optical lattice potential. The advantage of a lattice potential is the 
possibility of quenching the particle motion by tuning the hopping and 
thus vary the ratio of dipolar potential energy and motional kinetic 
energy in order to study transitions from quantum delocalized phases 
driven by kinetic motion towards the classical regime of large dipolar
interaction. For dipolar particles in 1D with dipole moments 
perpendicular to the 1D tube, a transition from a linear
configuration to a zig-zag phase has been predicted \cite{astrakharchik2008a,astrakharchik2008b} 
using both classical arguments and quantum Monte Carlo calculations and 
can be driven by tuning the transverse confining potential that keeps the
particles in a 1D geometry. In a more recent paper \cite{ruhman2012}
the same system was studied using Luttinger liquid techniques and 
the density matrix renormalization group (DMRG) and evidence for a
non-local string ordering was presented that could give rise to edge
states. In the DMRG this was done using a nearest-neighbour Hubbard-type
model that was expected to show similar physics to the true system of 
dipolar particles.

Here we consider a Hubbard model that includes the dipole-dipole
interaction for different orientations of the dipole moments of the 
particles with respect to the two-leg ladder geometry. By varying the
hopping parameters and the dipole strength, we look for zig-zag ordering
by using the string and Ising ordering parameters introduced in 
Ref.~\cite{ruhman2012} obtained through the Matrix Product States (MPS)
technique which is essentially equivalent to DMRG. 
This is done for different transverse hopping
parameters and different strengths of the dipolar interaction. 
Our results demonstrate that there is a rich phase
diagram in the system as function of the orientation of the 
dipole moment, and that the ordering changes character as
the angle is varied. Surprisingly, we find that there is 
even a critical angle at which the system becomes completely
disordered independently of the transverse hopping and dipolar
strength. This critical angle occurs in the regime where 
two dipoles on the same leg of the ladder will repel 
each other. However, it is only a few degrees above the 
angle at which the interaction vanishes along the legs, 
indicating that it is the hopping along the legs that destroys
ordering at the critical angle.

The paper is organized as follows. In Sec.~\ref{formal} we introduce
the formalism, the MPS technique we have employed, and the 
order parameters we study. Sec.~\ref{res} contains our results
discussing first the spatial and Fourier transforms of the ordering
for different angles, then we present phase diagrams for different
filling fractions, and finally we discuss the phase diagram for 
fixed filling fraction and dipolar strength as the orientation 
angle and transverse hopping parameter is varied. In Sec.~\ref{con}
we discuss results and conclusion, as well as the experimental 
implementation of the system we study and detection schemes. 
In the appendix we elaborate on the numerical details, discuss
finite size effects, and the truncation of the dipolar interaction
and its influence on the order parameters.

\begin{figure}[ht!]
\centering
\includegraphics[scale=0.33]{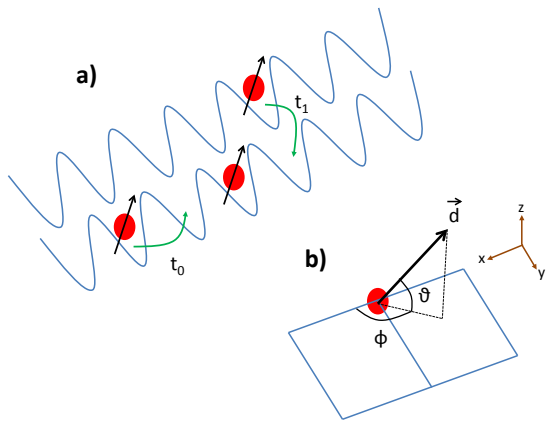}
\caption{Schematic visualization of the two-leg ladder setup with dipoles. a) The lattice
contains two legs and allows for hopping along legs with parallel hopping parameter $t_0$ and 
along the rungs with perpendicular hopping parameter $t_1$. The polar molecules are shown as filled
red circles in the lattice with an arrow that indicates that they have a dipole moment. b) Specification 
of orientation of the dipoles with respect to the ladder. The angle $\phi$ is measured from the $x$-axis
while $\theta$ is the angle out of the plane of the ladder. On the right we show the coordinate system we
will use with legs along the $x$-axis and rungs along the $y$-axis.}
\label{setup}
\end{figure}

\section{Formalism}\label{formal}
We study a two-leg ladder, i.e. a two parallel one-dimensional lattices
that hold polar molecules which are allowed to hop both along the 
two legs (parallel hopping) and across the rung (perpendicular hopping). 
The setup is shown in Fig.~\ref{setup}. The lattice constant, $a$, is
the same along the legs and across the rungs. The interaction of two 
polar molecules depends on their relative distance, $r$, through
\begin{align}
V(\bm r)=\frac{D^2}{r^3}\left(1-3\cos^2\alpha\right),
\end{align}
where $D^2=d^2/4\pi\epsilon_0$ with $d$ the dipole moment and $\epsilon_0$ 
the vacuum permittivity. The angle $\alpha$ is defined through the 
distance and dipole moment vector, $\bm d$, via $\cos\alpha=\bm r\cdot\bm d/rd$.
Refering to Fig.~\ref{setup}, we parametrize $\bm d$ by
\begin{align}
\bm d=d(\cos\theta\cos\phi,\cos\theta\sin\phi,\sin\theta).
\end{align}

The two-leg ladder is described by a tight-binding model with Hamiltonian
\begin{align} \label{Hamiltonian}
H=&-{\sum_{i,\sigma}}'\left(t_0 a_{i,\sigma}^{\dagger}a_{i-1,\sigma}
+t_1 a_{i,\sigma}^{\dagger}a_{i,-\sigma}+\textrm{h.c.}\right)&\nonumber\\
&+\sum_{i,\sigma,k,\sigma'}V_{i,\sigma,k,\sigma'}n_{i,\sigma}n_{k,\sigma'},&
\end{align}
where $a_{i,\sigma}^{\dagger}$ and $a_{i,\sigma}$ are bosonic
operators that create and annihilate particles on the site with leg
position $i$ (in units of the lattice constant $a$) and on the side of
the rung indexed by $\sigma=\pm 1$. The $\sum'$ indicates that only
nearest-neighbour hopping terms are included. The number operator is
defined in the usual way,
$n_{i,\sigma}=a^{\dagger}_{i,\sigma}a_{i,\sigma}$.  The dipole-dipole
interaction term may now be written \cite{zinner2011-1D}
\begin{align}\label{dipint}
V_{i,\sigma,k,\sigma'}=&\frac{D^2}{a^3}\left[
\frac{1}{((i-k)^2+\delta_{\sigma,-\sigma'})^{3/2}}\right.&\nonumber\\
&\left.-\frac{3\cos^2\theta\left(\cos\phi(i-k)+\sin\phi\delta_{\sigma,-\sigma'}\right)^2}{((i-k)^2+\delta_{\sigma,-\sigma'})^{5/2}}\right],&
\end{align}
where $\delta_{\sigma,-\sigma'}$ is zero for particles on the same leg and one for particles on opposite legs.
In the numerical calculations we need to truncate the dipolar interaction which we will do 
at the next-next-nearest neighbour level (three sites removed). In appendix~\ref{range}
we discuss the effects of truncation, and in particular the fact that previous studies that
use truncation at nearest neighbour interactions (as in the so-called extended Bose-Hubbard models)
may not be able to capture the full correlations induced by the long-range dipolar interactions.
From now on we will use $t_0$ as the unit of energy. This means that $V_d=D^2/a^3/t_0$ is the 
dimensionless measure of the dipolar interaction strength, while $t_1/t_0$ measures the 
perpendicular hopping normalized to the parallel one. 

We employ particle number-conserving matrix product states with each rung of the ladder as the site Hilbert spaces. 
In order to find the ground state of the Hamiltonian (\ref{Hamiltonian}), we use repeated sweeps of local energy minimizations until the energy has converged to within a relative tolerance \cite{verstrate2008} which we set to $10^{-6}$.
The Hamiltonian is represented by a matrix product operator whose bond-dimension depends linearly on the dipole-interaction truncation length.
In order to keep the lattice site Hilbert space dimension small, we have assumed hard-core bosons, 
which can be achieved experimentally by tuning the on-site 
interaction $U n_{i\sigma} (n_{i\sigma} - 1) / 2$ large using, e.g., a Feshbach resonance.

\begin{figure*}[ht!]
\centering
\includegraphics[scale=0.6]{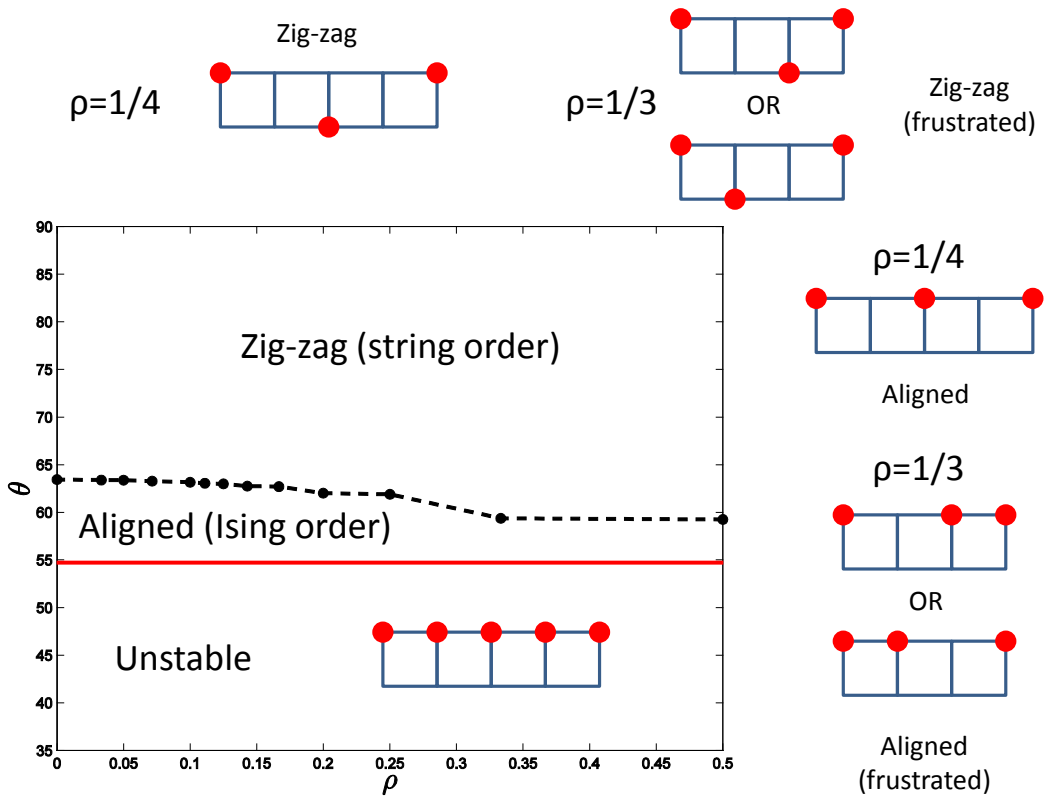}
\caption{Classical phase diagram ($t_0=t_1=0$) as function of density $\rho$ and angle $\theta$. The various
classical configurations are illustrated at the top and right-hand side for $\rho=1/4$ and $\rho=1/3$. The 
latter case has frustration as shown in the drawings. This is generally true for $\rho=1/p$ with $p$ an odd
integer. Notice that
below the solid horizontal line ($\theta_M=\cos^{-1}(1/\sqrt{3})$) there is attraction along the legs and the 
particles will cluster on one side and increase the density. We consider hard-core bosons here so that we still
have at most one particle per site.}
\label{class}
\end{figure*}

The Ising and string ordering that we compute is defined through the 
functions \cite{ruhman2012}
\begin{eqnarray}
&C_I(r_i-r_j)=\langle\sigma_i\sigma_j\rangle,&\label{ising}\\
&C_\textrm{string}(r_i-r_j)=\langle \sigma_i \exp\left(i\pi\sum_{l=j}^{i}n_i\right)\sigma_j\rangle,&\label{string}
\end{eqnarray}
where $\sigma_i=n_{i,1}-n_{i,2}$. Here $n_{i,1}$ and $n_{i,2}$ are the density operators for 
site $i$ on legs 1 and 2 respectively. Notice that we have left out the factor 
$\cos\left(\pi\rho(r_i-r_j)\right)$ in the Ising order as compared to Ref.~\onlinecite{ruhman2012}, where
$\rho$ is the density of particles. Using this defintion we may then do a general Fourier transform
of the Ising order (and string order as well) making it easier to identify periodicity even when 
it is not directly related to $\rho$. All Fourier transforms below are with respect to the 
basis function $\cos(\nu\pi(r_i-r_j))$ where $\nu$ has units of $1/a$. We denote the 
Fourier transform by a tilde, i.e. $\tilde C_I(\nu)$ and $\tilde C_\textrm{string}(\nu)$. 
A peak at $\nu=\rho$ therefore recovers the same
ordering as the one discussed in Ref.~\cite{ruhman2012}.
We will calculate these functions starting from the 
center of our system and moving toward the edge in order to minimize finite-size effects.
In appendix~\ref{size} we include a quantitative discussion of finite-size effects in 
our setup.

\section{Results}\label{res}
The following subsections presents and discusses our results for selected parameters
of interest. This includes different dipolar strengths, transverse hoppings, orientation
angles, and filling fractions. For simplicity we have used $\phi=0$ throughout the paper, 
postponing a discussion of non-zero $\phi$ to future studies.

\subsection{Classical phase diagram}
In Fig.~\ref{class} we show a classical phase diagram calculated by assuming that 
the hoppings are zero ($t_0=t_1=0$). As function of $\rho$ and $\theta$ we find
three different regimes. A zig-zag phase is present for large $\theta\gtrsim \pi/3$, 
while for $\pi/3\lesssim \theta>\theta_M=\cos^{-1}(1/\sqrt{3})$ an aligned phase
occurs. The configurations of these phases are shown in Fig.~\ref{class} for
the case of $\rho=1/3$ and $\rho=1/4$. In the case with $\rho=1/p$ where $p$ 
is an odd integer, there is a frustrated configuration with an energy minimum 
that is degenerate as illustrated for $p=3$ in Fig.~\ref{class}. The critical 
angle between the aligned and the zig-zag phases depends only weakly on $\rho$. 
For $\rho=1/N$ and in the limit where $N\to\infty$, one can show that the critical angle goes to 
$\theta=\cos^{-1}(1/\sqrt{5})$ by comparing the energies of the two configurations.
This is consistent with our numerical energy minimization. For $\theta<\theta_M$, the
interaction on a single leg is attractive and
the minimum has all particles on the same leg and as close as possible, i.e. 
there will be a change of density at this point. Notice that we use hard-core
bosons and thus can have at most one particle per site. 
In the numerical calculations with non-zero hopping below we will see that
the zig-zag phase is associated with the string order, while the aligned
phase is associated with the Ising order. The classical phase diagram is therefore
very instructive in understanding the behavior of the full quantum problem.

\subsection{Perpendicular dipoles}
We first focus on the case where the dipoles are perpendicular to the plane containing
the two-leg ladder, i.e. $\theta=\pi/2$ and $\phi=0$. This means that the dipole-dipole
interactions are purely repulsive. This case was discussed recently in Ref.~\cite{ruhman2012}
where numerical DMRG calculations are discussed for two choices of parameters in a model 
for the dipolar system including nearest neighbour interactions only and different interactions along the 
ladders as compared to along the rungs. Here we consider the full dipolar interaction
given in Eq.~\eqref{dipint} (with truncation at next-next-nearest neighbour 
for convergence, see Appendix~\ref{range}).

\begin{figure}[t!]
\centering
\includegraphics[scale=0.42]{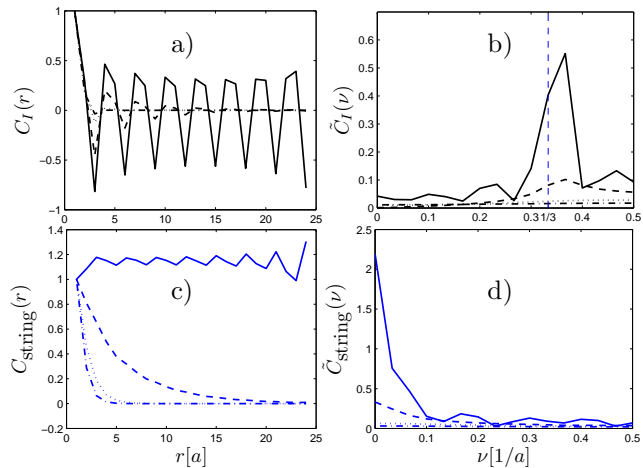}
\caption{String and Ising order parameters for $L=48$ and $\rho=1/3$.
The dipolar angles are $\theta=\pi/2$ and $\phi=0$, while the
interaction strength is $V_d=1.0$. The transverse hopping is $t_1=0.1$ (solid), 
$t_1=0.2$ (dashed), $t_1=0.5$ (dotted), and $t_1=1.0$ (dash-dotted).
Panel a) shows the Ising order measured from the center of the system and towards the edge and 
b) its Fourier transform. Panels c) and d) are similar plots for the string order.
The vertical dashed line in panel b) indicates $\nu=1/3$.}
\label{perpfig}
\end{figure}

In Fig.~\ref{perpfig} we show results of a calculations on a lattice of size $L=48$ with 
filling $\rho=1/3$ and dipolar interaction $V_d=1.0$. The perpendicular hopping, $t_1$, 
is varied from 0.1 to 1.0 in four steps. For $t_1=0.1$ we clearly see the co-existence 
of both string and Ising orders, while for $t_1\geq 0.2$ both orders show exponential 
decay. In the ordered regime, the string correlations are almost constant but with a 
small oscillatory component. Interestingly, this oscillatory behavior actually persists
for other system sizes (see Appendix~\ref{size}) with matching maxima and minima 
(except at the boundaries). While tempting, interpreting these oscillations as being
due to finite-size seems not the case. In Fig.~\ref{perpfig} 
we also see a clear boundary effect in the string order
at the edge of the system as expected. As we discuss in Appendix~\ref{size} we 
expect only minute quantitative changes in our results for larger system sizes.

\begin{figure}[t!]
\centering
\includegraphics[scale=0.40]{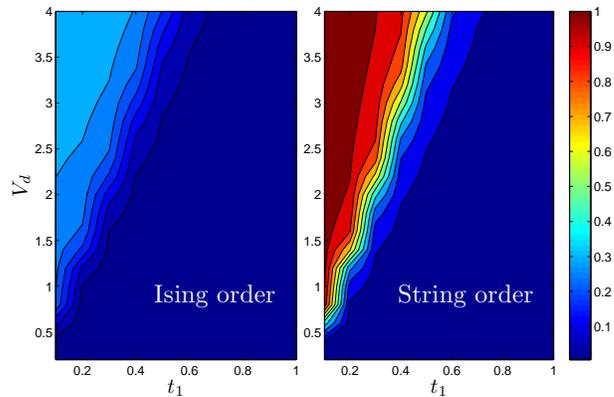}
\caption{Phase diagrams for the Ising (left) and string (right) orders as a function of
$V_d$ and $t_1$ for filling factor $\rho=1/3$ and perpendicular dipoles, $\theta=\pi/2$. 
The orders have been computed by taking the absolute value of the largest
Fourier component of $\tilde C_I(\nu)$ and $\tilde C_\textrm{string}(\nu)$ respectively. 
The values have been normalized to the largest value over the entire
region of parameter space.}
\label{phaseperp}
\end{figure}

For larger values of $V_d$ we find that the orders persist to larger $t_1$ and the 
relevant parameter is $V_d/t_1$ (not shown here). 
This can be quantified by working out the phase diagram
for the string and Ising orders as function of $V_d$ and $t_1$. In Fig.~\ref{phaseperp} 
we present the phase diagram which has been computed by using the largest Fourier 
component for each value of $V_d$ and $t_1$. The filling factor is $\rho=1/3$. 
The order parameters are both non-zero 
for the larger part of the phase diagram. However, for low perpendicular hopping
we see that the string order extends to smaller values of $V_d$, implying that Ising
order vanishes slightly faster. 

\begin{figure}[t!]
\centering
\includegraphics[scale=0.39]{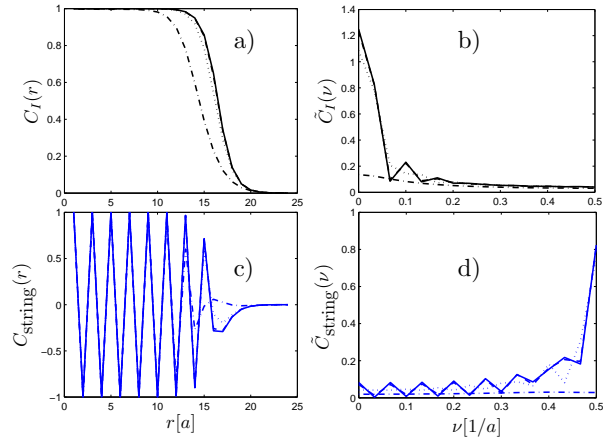}
\caption{String and Ising order parameters for $L=48$ and $\rho=1/3$. The dipolar angles
are $\theta=0$ and $\phi=0$, while the 
interaction strength is $V_d=1.0$. The transverse hopping is $t_1=0.1$ (solid), 
$t_1=0.2$ (dashed), $t_1=0.5$ (dotted), and $t_1=1.0$ (dash-dotted).
a) shows the Ising order measured from the center of the system and towards the edge and 
b) its Fourier transform. c) and d) are similar plots for the string order.}
\label{parfig}
\end{figure}

\subsection{Transition to parallel dipoles}
If the dipoles are instead oriented parallel to the legs of the ladder, we have 
purely attractive interactions on the legs, while the interaction across the rungs
remain repulsive at short distance and then becomes attractive for larger distances
between two sites. In Fig.~\ref{parfig} we show the profiles and Fourier transforms
for parallel dipoles ($\theta=\phi=0$) with all other parameters the same as in 
the perpendicular case in Fig.~\ref{perpfig}. The first observation is that the 
roles of Ising and string orders are reversed with respect to constant and 
oscillatory behaviors. 
This is consistent with the strong attraction along the legs making it favourable 
for the particles to cluster on one side of the system (which side is chosen is a
matter of small breaking of symmetry in the initial condition of the calculation). 
We also notice that the Ising order stays completely constant in the bulk region, 
avoiding the oscillatory component that is seen in the perpendicular case
in Fig.~\ref{perpfig} in the string ordered phase.

\begin{figure}[t!]
\centering
\includegraphics[scale=0.39]{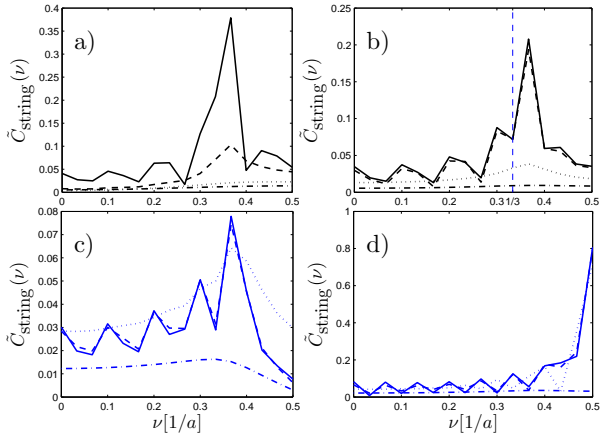}
\caption{String order parameter for $L=48$ and $\rho=1/3$. The dipolar angles
are $\theta=\pi/4$ and $\phi=0$. The transverse hopping is $t_1=0.05$ (solid), 
$t_1=0.2$ (dashed), $t_1=0.5$ (dotted), and $t_1=1.0$ (dash-dotted). The
interaction strength is
a) $V_d=1.0$, b) $V_d=2.2$, c) $V_d=3.0$, and d) $V_d=3.8$. Note the different
vertical scales on the four panels.
The vertical dashed line in panel b) indicates $\nu=1/3$.}
\label{45fig}
\end{figure}

A striking difference to the perpendicular case 
is that the oscillation around zero in the string order is not around $\nu=1/3$
but rather $\nu=1/2$. This again is a result of the attraction on the legs which means
that the cluster of particles is uniformly distributed along a leg, thus changing the
sign of the string order for each site. For smaller $V_d$ one can recover the peak at $\nu=1/3$
but only for small $V_d\sim 0.1$ and for correspondingly small $t_1$ (in order to see ordering
at all). This means that one can in principle choose a small $t_1$ value and then 
tune $V_d$ to follow the transition from string ordering with two different characteristic
periods. In contrast, for perpendicular dipoles the interactions are purely repulsive
and the zig-zag phase with $\nu=1/3$ will be generated for any $V_d$ at sufficiently 
small $t_1$.

The above considerations begs the question as to what happens at some intermediate 
angle between the extreme case of perpendicular and parallel dipoles. In Fig.~\ref{45fig}
we therefore show results for $\theta=\pi/4$ and $\phi=0$ for the string order at 
different values of $V_d$. In this case the attraction along the legs is present
but much smaller than for parallel dipoles, while the repulsion across the rung
is the same as both parallel and perpendicular orientations. The figure demonstrates
a transition from $\nu=1/3$ to $\nu=1/2$ as $V_d$ increases as one goes from a 
zig-zag configuration driven by repulsion to a clusterization on one leg driven
by the attraction. This is consistent with the classical behavior discussed 
above.
The Fourier coefficients spread out and their magnitudes
decrease (note the different vertical scales in Fig.~\ref{45fig}) as we transition
to the cluster state at large $V_d$. In our numerical calculations the cluster
state is only stabilized by the hard-core nature of the bosons. If this 
condition is removed the system will have a negative compressibility.
This also suggests that the transition to the clustered state could be
first order transition since the density changes as we cross the 
critical angle towards the attractive regime at small $\theta$.
This is an example of the competition 
between different signs of the dipolar interaction and its influence on the 
non-local string order. This should have a clear signal in experimental setups
that seeks to measure string correlations.

\subsection{Phase diagrams}
We now consider the different phase diagrams that are possible to explore through a 
change of $V_d$, $t_1$, $\theta$, and the filling fraction. Note that the phase diagrams
that we plot here show the strength (absolute value of largest Fourier component) of
the Ising and string order (given by the correlation functions in Eqs.~\eqref{ising} and \eqref{string}).
As the two orders tend to be non-vanishing at the same time (but of different magnitude), 
we plot the two in separate graphs. The more traditional approach would be to plot 
where each are non-zero or vanishing in a single plot, but we feel that the present 
form contains more information.

The first case is 
perpendicular dipoles which is shown in Fig.~\ref{phaseperp} for $\rho=1/3$.
The left side gives
the Ising while the right side shows the string order (color coding on the far right).
Note that we have normalized the order parameters so that their largest value in the 
whole phase diagram is one. 
As $t_1$ increases, both orders eventually go to zero as expected. What is more interesting
is that they tend to be pinned to each other and disappear along roughly the same 
line in the phase diagram. In the ordered phase, however, the string order tends
to have a larger (relative) magnitude than the Ising order. This should be compared to the 
results presented in Ref.~\onlinecite{ruhman2012} where phases of vanishing Ising but
non-vanishing string was obtained. The model in Ref.~\onlinecite{ruhman2012} has large
repulsion along the rungs, uses only nearest neighbour interactions, 
and is thus different from the more realistic dipole-dipole interaction that we consider here.

\begin{figure}[t!]
\centering
\includegraphics[scale=0.40]{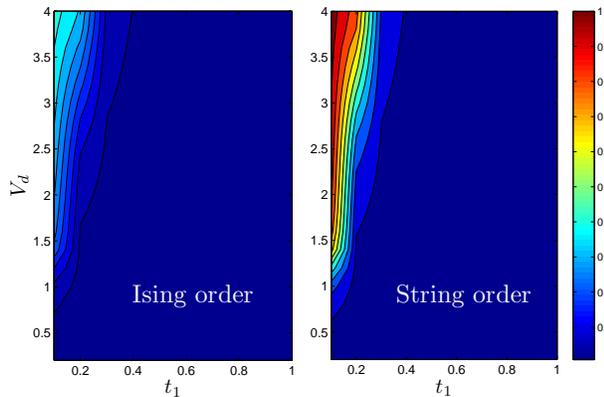}
\caption{Phase diagrams for the Ising (left) and string (right) orders as a function of
$V_d$ and $t_1$ for filling factor $\rho=1/4$ and perpendicular dipoles, $\theta=\pi/2$.}
\label{phasequarter}
\end{figure}

For lower filling fractions the orders are expected to be less robust as the 
transverse hopping is increased since correlations vanish faster at lower
filling. This can be seen in Fig.~\ref{phasequarter} which is also for the 
case of perpendicular dipoles, but this time with filling $\rho=1/4$. The 
picture is similar to that in Fig.~\ref{phaseperp}, but with a severely
shrunk ordered region. This is generally the case for small $\rho$.
The string and Ising orders are still pinned to
each other, with the string slightly stronger in the ordered region. For 
larger filling fractions (not shown here), we find larger ordered regions 
as expected. 

\begin{figure*}[ht!]
\centering
\includegraphics[scale=0.8]{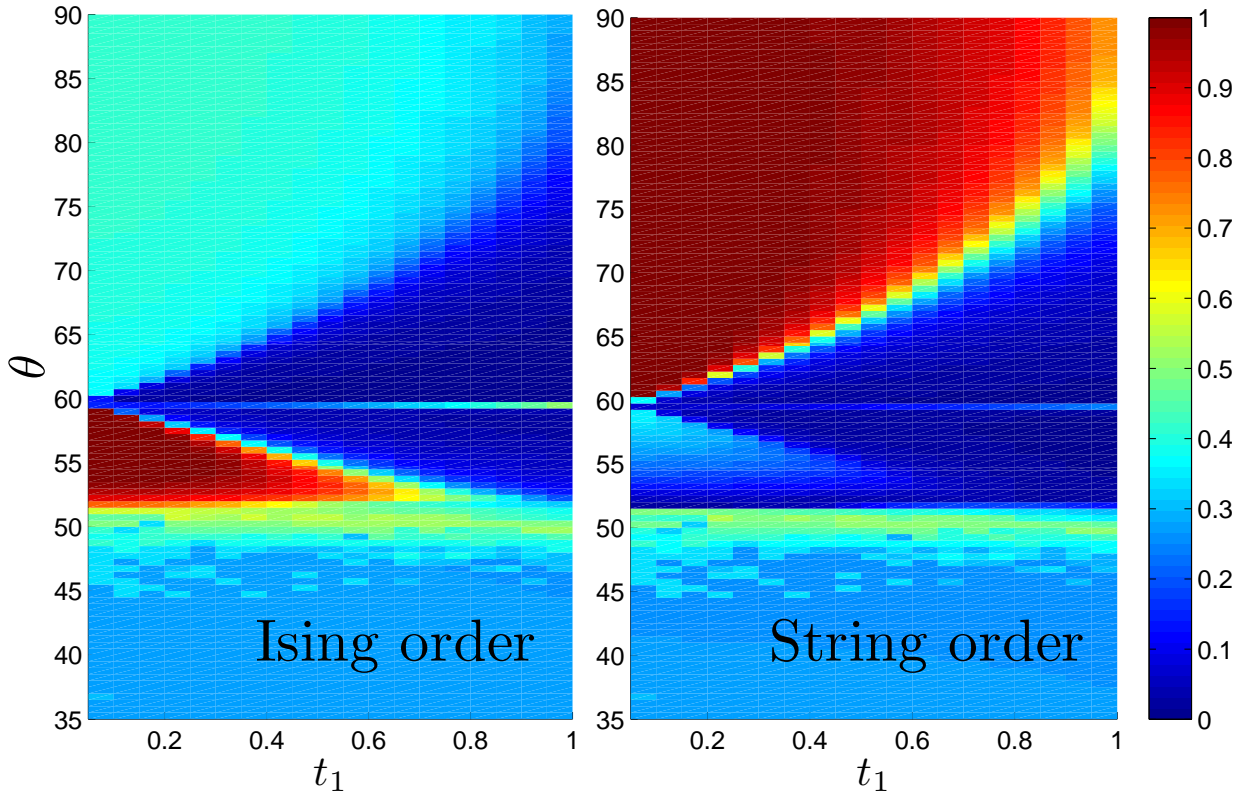}
\caption{Phase diagrams for the Ising (left) and string (right) orders as a function of
tilting angle $\theta$ (in degrees) and $t_1$ for filling factor $\rho=1/3$ and $\phi=0$. The 
dipolar strength is $V_d=10$.}
\label{phaseangle}
\end{figure*}

We now consider the phase diagram as a function of $\theta$ which is shown in Fig.~\ref{phaseangle}.
Our calculations show that a critical angle $\theta_0\sim \pi/3$ exists at which the 
system has neither Ising nor string order for any value of $t_1$. This critical angle seems to be 
independent of $\rho$ and $V_d$ according to numerical results. This is supported by the 
classical calculations shown in Fig.~\ref{class}. If we consider small transverse
hopping, $t_1\ll 1$, then $\theta_0$ is the point at which the system changes character from 
being Ising ordered at $\nu=0$ due to clustering on one leg for $\theta<\theta_0$ to being string ordered due to
the repulsively induced zig-zag behavior at $\nu=1/3$ for $\theta>\theta_0$. 
For smaller values of 
$V_d$ the maximum Fourier component will change its frequency $\nu$ (as seen in Fig.~\ref{45fig})
but the behavior is the same with a critical $\theta_0$ at which the order parameters vanish.
It is interesting to compare the angle $\theta_0$ to the so-called magic angle,  
$\theta_M=\textrm{arccos}(\tfrac{1}{\sqrt{3}})\sim 0.955 \sim 55^\circ$ \cite{lahaye2009,santos2010,deu2010,zinner2011-1D},
at which two dipoles with this orientation will be non-interacting. This is about 5 degrees smaller than
$\theta_0$. Since $\theta_0>\theta_M$ we see that it takes a finite amount of repulsive strength along
the legs of the ladder to go from Ising order to string ordering at very small $t_1$. This is 
in spite of the fact that $V_d$ is large compared to the hopping along the legs in the phase
diagram in Fig.~\ref{phaseangle}. The dipolar interactions also act along the diagonals of the 
ladder (also up to third removed neighbour site) and diminishes the repulsion from the longer
ranges (although it never switches sign). The local repulsion from the site directly across
on the other leg, however, appears to be insufficient for ordering for any $t_1$. Below angles
of $\theta\sim 50$ degrees both order parameters become frustrated as the interactions 
become dominantly attractive, as seen also in the classical phase diagram. Note 
that the quantum delocalization moves the critical angle to 50 degrees which is 
slightly below the classical value of $\theta_M=54.7$ degrees.
For small angles both order parameters settle on a (small) non-zero value. 
However, in this part of the phase diagram, the system is effectively 
attractive and will be unstable as discussed above.

The diagram in Fig.~\ref{phaseangle} was obtained using a large value $V_d=10$. 
Smaller values of $V_d$ give qualitatively
the same behavior and the only significant difference is a shrinking of the ordered 
region to smaller values of $t_1$. This is consistent with the fact that the 
classical transition from zig-zag to aligned does not depend on the size of $V_d$.
Also, we have used a filling of $\rho=1/3$. Smaller
fillings will shrink the ordered region, while at $\rho=1/2$ the region of order is 
maximized such that the magnitude of the order is larger everywhere expect for the 
wedge termination at the critical angle $\theta_0$ that is still the same as for $\rho=1/3$
from Fig.~\ref{phaseangle}.

An interesting question beyond the scope of the current paper concerns the low-energy
theory around the critical angle where both orders vanish. Similar geometries without
the in-tube lattice have been studied using Luttinger liquid theory 
\cite{citro2007,citro2008,kollath2008,huang2009,dalmonte2010,fellows2011,leche2012}.
In particular, Ref.~\onlinecite{chang2009} has presented a phase diagram using 
a Luttinger approach for $\phi=0$ and $\theta$ around our critical value for the 
zig-zag to aligned transition, although with no intertube hopping term and using 
fermionic dipolar particles. Since we use hard-core bosons, this should nevertheless
be comparable. For large angles close to $\theta=\pi/2$, a density-wave state is 
found which is consistent with our findings. Around the critical angle, there
could be other phases appearing \cite{chang2009,huang2009,leche2012}, and with
a lattice in the tube a number of crystalline phases have been discussed 
\cite{dalmonte2010,dalmonte2011,bauer2012}. It would be interesting
to combine bosonization and mean-field theory as described in Ref.~\onlinecite{fellows2011}
with the in-tube lattice of the present setup (possibly along the lines of 
Ref.~\onlinecite{dalmonte2010}).

\section{Conclusions}\label{con}
We studied a two-leg ladder containing dipolar particles as function of 
interaction strength, transverse hopping, and orientation angle of the 
dipole moments with respect to the ladder plane in order to investigate
the presence of the quantum zig-zag phase. Within a Hubbard-type model 
with long-range dipole-dipole interaction, we computed Ising and string
correlation that display ordered phases for repulsive dipolar interactions
and low transverse hopping. Two different kinds of string ordering was
obtained for filling fraction of one-third with different characteristic 
frequencies and a transition between these takes place as the orientation 
of the dipoles is changed from being parallel to perpendicular to the ladder
plane. Furthermore, we find that there is a critical angle of about
60 degrees at which neither Ising nor string ordering survives at finite
transverse hopping. Our numerical calculations indicate that this critical angle
is insensitive to filling fraction and dipolar strength.

An interesting question concerns the order of the transitions in our phase diagrams
as function of dipolar strength for fixed angle or vice versa, and for different 
values of the transverse hopping. We have seen in Fig.~\ref{45fig} that as the 
angle is varied the Fourier components will diminish, flatten out, and re-emerge
with a peak at a different frequency as we tune across the transition of the 
different kinds of string order. This seems to be a smooth behavior and not a
first order transition. However, as we have discussed above this is driven 
by attractive interactions and there may thus indeed be a change in density 
once the interaction is strong enough to compensate the kinetic energy in the
hopping terms. This would be a sudden change in density and could indeed be
a first order transition taking place for a sufficiently small and fixed $\theta$ 
as function of $V_d/t_1$. 
The phases for fixed dipolar strength, $V_d$, as function of the 
angle, $\theta$, in Fig.~\ref{phaseangle} have a rather abrupt behavior 
around the critical angle where the both Ising and string orders vanish. This 
is reminiscent of the density-wave state discussed recently in Ref.~\onlinecite{gammelmark2013}.
This could indicate a first order transition. On the other hand, as we saw in the 
classical phase diagram the zig-zag to aligned transition will not change the 
total density of the system but merely the density on individual legs of the ladder. 
This would suggest that the transition could be smooth both for vanishing 
and finite transverse hopping, $t_1$. This would imply that string and Ising
orders do not compete in the effectively repulsive regime, $\theta>\theta_M=\textrm{cos}^{-1}(1/\sqrt{3})$.
Further investigation seems necessary to settle these issues.

In order to implement the model we have studied here experimentally, one can use
heteronuclear molecules with dipolar moments that can be controlled by externally
applied fields
\cite{ospelkaus2008,ni2008,deiglmayr2008,lang2008,ospelkaus2010,ni2010,shuman2010,miranda2011,chotia2012}. 
Alternatively, one can also use neutral atoms with sizable magnetic dipole moments
\cite{griesmeier2005,lahaye2007,lu2010,lu2011,lu2012} or Rydberg atoms \cite{gallagher2008,saffman2010,comparat2010}.
Measuring the bulk ordering such as Ising and string orders seems very realistic given
current experimental developments using single-site addressing \cite{bakr2009,sherson2010}. 
In particular, string order in Mott insulators
have been probed in one dimension \cite{endres2011} and crystal structure has been
probed in a Rydberg gas in two dimensions \cite{schauss2012}. Even dynamics has been 
accessible in recent experiments \cite{cheneau2012}. The two-leg ladder geometry
can be realized using a superlattice in the transverse confinement direction which 
should be compatible with the observation techniques.

\begin{acknowledgments} 
We thank Erez Berg for illuminating discussions and Klaus M{\o}lmer for his 
continuous encouragement
and support.
N.T.Z. is supported by the Danish Council for Independent Research - Natural Sciences, 
while S.G. acknowledges support from the Villum Kann Rasmussen foundation.
\end{acknowledgments}

\appendix

\section{Numerical details}\label{app-num}
\begin{figure}[htb]
  \centering
  \includegraphics[width=\columnwidth]{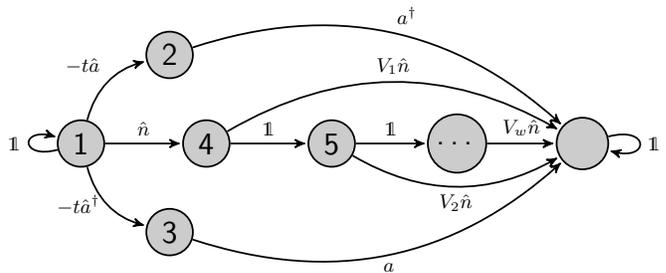}
  \caption{Illustration of a long-range interaction MPO for a single leg.}
  \label{fig:MPO}
\end{figure}

The numerical method used here is well-documented 
and compared to other methods in wide use (such as the 
density matrix renormalization group method) in the 
papers by Verstreate {\it et al.} \cite{verstrate2008}
and by Crosswhite and Bacon \cite{crosswhite2008}. We 
will therefore not enter a discussion of the full numerical details
but only sketch a few of the essential ideas.
The long-range Hamiltonian is represented by a single matrix 
product operator (MPO) in our calculations.
A simple way of constructing the matrices of the MPO is to utilize 
the connection with finite automata or hidden Markov models 
\cite{verstrate2008,crosswhite2008}, where the matrices represents 
the action of a finite automaton or the transition rules of a hidden 
Markov model where the emitted symbols are exactly the strings of 
operators whose tensor product occurs in the Hamiltonian Eq.~\eqref{Hamiltonian}.
A hidden Markov model for a single leg of dipoles with long-range 
interactions is illustrated in Fig.~\ref{fig:MPO} and the generalization 
to multiple legs is straight forward.
By using an MPO-formulation of the Hamiltonian it is straight forward 
to evaluate the variance of the energy of the calculated state which 
provides an independent bound on the accuracy of the ground-state energy.

\begin{figure}[ht!]
\centering
\includegraphics[scale=0.42]{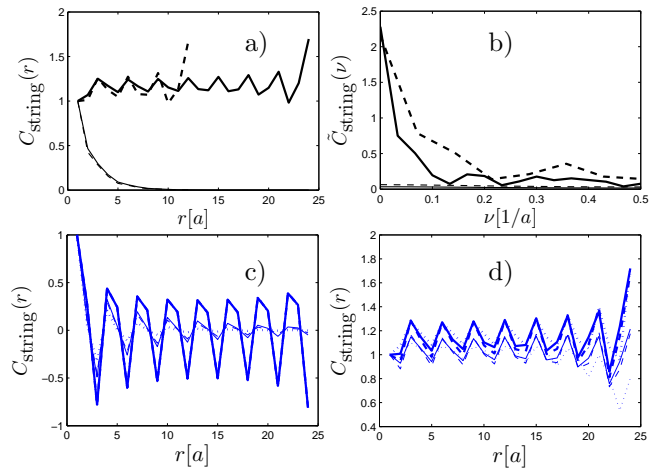}
\caption{Exploration of finite size and dipolar interaction range
truncation effects on the string order for $\phi=0$ 
and $\rho=1/3$. a) String order parameter profile for $V_d=4.0$ and $\theta=\pi/2$ 
with $L=48$ ($N=32$) 
as solid line and $L=24$ ($N=16$) as the dashed line. The upper curves have $t_1=0.1$ 
and the lower ones have $t_1=1.0$. b) Fourier transform of the  profiles in a). 
c) $V_d=10$ and $\theta=0.96$ (55 degrees) for truncation of the dipolar interaction
at nearest (dotted line), next-nearest (dashed line), and next-next-nearest neighbour (solid line).
Profiles for $t_1=0.05$ and $t_1=0.5$ are shown. d) Same as in c) for $\theta=1.22$ (70 degrees).}
\label{fin-range}
\end{figure}

\section{Finite size effects}\label{size}
To investigate potential effects of the finite system sizes that we work with, we have checked that 
our results are insensitive to increases in system size while keeping the filling fraction, $\rho$, 
the same. In Fig.~\ref{fin-range} panels a) and b) we show an example of the behavior of the string
correlation order parameter as the size is increased from $L=24$ (dashed lines) to $N=48$ (solid lines) 
while keeping $\rho=1/3$. 
The results shown have $V_d=4.0$, $\phi=0$, and $\theta=\pi/2$, i.e. perpendicular dipoles. In 
Fig.~\ref{fin-range}a) the upper curves that oscillate around a finite value are for $t_1=0.1$ while
the decaying profiles in the lower part have $t_1=1.0$. The former are in the string ordered phase
while the latter are in the string disordered phase. The curves show a very good resemblance of 
profiles for both the average values and the oscillatory peaks irrespective of system size. This 
shows that our results do not suffer from sizable finite size effects. For comparison, we show 
in Fig.~\ref{fin-range}b) the Fourier transform of the profiles in Fig.~\ref{fin-range}a). 
Here we do see some differences at larger frequency values for the $t_1=0.1$ case, but only
of sub-leading nature in comparison to the main peak at zero frequency which is properly reproduced.

\section{Dipolar interaction range}\label{range}
Another issue with the implementation of dipolar interactions in the matrix product state procedure
is the necessary truncation at some finite neighbour distance. In order to gauge whether this 
truncation posed can potentially change any of the results presented in this paper, we have 
studied the behavior of the system as the number of neighbour sites over which the dipolar 
force acts is varied. In Fig.~\ref{fin-range}c) and d), we show the string order parameter
for $L=48$, $\rho=1/3$, $V_d=10$, and for two different values of $t_1=0.05$ (thick lines) 
and $t_1=0.5$ (thin lines). The angles are $\phi=0$ and $\theta=0.96$ (55 degrees) which 
is in the string disordered phase (thus the oscillatory pattern around zero). The truncation 
has been done at nearest (dotted), next-nearest (dashed), and next-next-nearest (solid) neighbour
sites. For the lower $t_1=0.05$ the three curves are virtually the same and no truncation 
effects can be seen. However, as we increase to $t_1=0.5$, we see that the nearest neighbour
truncation differs from the others, while next and next-next values are very close. This 
shows that truncation at next-nearest neighbour is adequate for our purposes, while using
only nearest neighbour dipolar interactions would miss some long-range correlations in the
system as we see by the faster decay in that case in Fig.~\ref{fin-range}c). The same
story plays out in Fig.~\ref{fin-range}d) where we have used $\theta=1.22$ (70 degrees)
and are thus in the string ordered regime of the phase diagram. Again we see that truncation
at only nearest neighbour tends to miss correlations at longer range for larger $t_1$.
The results presented in the main text have all be computed using the next-next-nearest 
neighbour truncation.

\end{document}